\documentclass[11pt]{article}
\pdfoutput=1
\usepackage[american]{babel}
\usepackage{fullpage}
\usepackage{amsmath}
\usepackage{hyperref}
\usepackage{graphicx}
\usepackage{amssymb}

\newcommand{\heading}[1]{\begin{center} {#1} \end{center}}

\newcommand{\spc}[1]{\hspace{#1cm}}
\newcommand{\sspc}{\spc{0.25}}
\newcommand{\lspc}{\spc{0.5}}
\newcommand{\shortcomma}{\spc{0.15} , \spc{0.35}}
\newcommand{\longcomma}{\spc{0.3} , \spc{0.7}}
\newcommand{\psik}{\spc{0.3} ,}
\newcommand{\nekuda}{\spc{0.3} .}
\newcommand{\nn}{\nonumber}

\newcommand{\Me}{M_E}
\newcommand{\Mnu}{M_\nu}
\newcommand{\rr}{r_{23}}

\newcommand{\tot}{\theta_{12}}
\newcommand{\ttth}{\theta_{23}}
\newcommand{\toth}{\theta_{13}}
\newcommand{\ph}{\varphi}
\newcommand{\cph}{\cos\ph}
\newcommand{\VEVph}{\langle\ph\rangle}
\newcommand{\e}{\varepsilon}
\newcommand{\Dt}{\Delta m_{21}^2}
\newcommand{\Dth}{\Delta m_{31}^2}
\newcommand{\Dsol}{\Delta m_{\rm sol}^2}
\newcommand{\Datm}{\Delta m_{\rm atm}^2}

\begin{document}

\heading{\LARGE \bf Fine tuning in an $A_4$-based Tri-Bimaximal neutrino-mixing model}
\vspace{0.3cm}

\renewcommand{\thefootnote}{\fnsymbol{footnote}}
\heading{\Large Shahar Amitai\footnote{shahar.amitai@weizmann.ac.il}}
\renewcommand{\thefootnote}{\arabic{footnote}}
\addtocounter{footnote}{-1}

\heading{\large \emph{Department of Particle Physics and Astrophysics, \\
Weizmann Institute of Science, Rehovot 76100, Israel}}
\vspace{0.5cm}

\begin{abstract}
The $A_4$ group stands in the basis of many models that predict Tri-Bimaximal neutrino mixing at leading order. We study the Altarelli-Feruglio $A_4$ symmetry model and show that in order to produce as small value of $\rr \equiv \Dt / |\Dth|$ as measured, it requires fine tuning. This observation is important for an evaluation of the model, since the problem it is trying to solve in the first place is the tuning of the three mixing angles. We get the required level of fine tuning for the model in both its basic form and its seesaw realization.
\end{abstract}
\vspace{0.3cm}

\section{Introduction} \label{sec:intro}

During recent years, increasingly accurate measurements of neutrino mass and mixing parameters have been made. A joint analysis \cite{Fogli:2008} of the KamLAND, SNO-III and MINOS experiments \cite{Abe:2008, Aharmim:2008, Adamson:2008}, made in 2008, resulted in values for the lepton mixing angles that are compatible with the Tri-Bimaximal (TB) values
\begin{align}
\sin \tot = \frac{1}{\sqrt{3}} \longcomma \sin \ttth = \frac{1}{\sqrt{2}} \longcomma \sin \toth = 0 \psik
\end{align}
first proposed in ref. \cite{Harrison:2002}, by better than $1.6 \sigma$. This has been the case since neutrino oscillations were first observed by the Super-Kamiokande experiment \cite{Fukuda:1998}, a decade earlier. For this reason, a lot of effort was put in developing models that predict TB mixing for the lepton sector, or a slight deviation from it. It has been found that a broken flavor symmetry based on the discrete group $A_4$ appears to be particularly suitable to reproduce this specific mixing pattern in leading order \cite{Ma:2001}.

Before we proceed to explain the reasoning of using $A_4$ as a source of TB mixing we should point out that, more recently, deviations from TB mixing that are quite substantial have been established. In particular, $\sin \toth \neq 0$ has been measured \cite{An:2012, Ahn:2012, Abe:2011a, Abe:2011b, Adamson:2011}:
\begin{align} \label{eq:uethree}
\sin \toth \simeq 0.15 \nekuda
\end{align}
(In addition, $\sin \ttth$ deviates from Bimaximal mixing by about 0.08, and $\sin \tot$ deviates from Trimaximal mixing by about 0.03.) This large deviation calls into question the idea that the approximately TB mixing is a result of a an approximate symmetry in nature, rather than an accident. More concretely, if a model that explains (\ref{eq:uethree}) as a consequence of an approximate symmetry requires fine tuning at a level stronger than $0.15$, the motivation for such a model becomes questionable.

Let us start with a brief introduction to the $A_4$ group \cite{Altarelli:2010}. $A_4$ is the group of even permutations of 4 objects. It consists of 12 elements, which are divided to four equivalence classes. $A_4$ can be generated by any two basic permutation $S$ and $T$ that satisfy
\begin{align}
S^2 = T^3 = (ST)^3 = 1 \nekuda
\end{align}
$A_4$ has four irreps (for four equivalence classes). Three of them are one-dimensional and one is three-dimensional. We can use $S$ and $T$ to describe the irreps. The three 1D irreps are obtained by
\begin{align}
1:   \lspc S &= 1 \sspc T = 1              \nn \\
1':  \lspc S &= 1 \sspc T = \omega^2           \\
1'': \lspc S &= 1 \sspc T = \omega   \psik \nn
\end{align}
and the 3D irrep is obtained by
\begin{align} \label{eq:irrep}
3: \lspc S = \frac{1}{3} \left( \begin{array}{ccc} -1  & 2  &  2 \\
                                           2  & -1 &  2 \\
                                           2  & 2  & -1 \end{array} \right) \sspc
T = \left( \begin{array}{ccc} 1 &  0       &  0     \\
                              0 & \omega^2 &  0     \\
                              0 &  0       & \omega \end{array} \right) \lspc \left( \omega \equiv e^{2\pi i/3} \right) \nekuda
\end{align}

$S$ and $T$ can help us to identify two subgroups of $A_4$. The subgroup $G_S$ ($G_T$) is the set of elements generated only by $S$ ($T$). Since $S^2 = T^3 = 1$, $G_S$ and $G_T$ are isomorphic to $Z_2$ and $Z_3$, respectively. $A_4$ can be broken down to each one of these subgroups by the VEV of a triplet $\ph = (\ph_1, \ph_2, \ph_3)$:
\begin{itemize}
\item A VEV of the form $\VEVph_S \equiv (1,1,1)$ breaks $A_4$ down to $G_S$. This is because $\VEVph_S$ is invariant under $S$, but not under $T$.
\item A VEV of the form $\VEVph_T \equiv (1,0,0)$ breaks $A_4$ down to $G_T$. This is because $\VEVph_T$ is invariant under $T$, but not under $S$.
\end{itemize}

We next explore the implications of TB mixing on the mass matrix. In the flavor basis (where the charged lepton mass matrix $\Me$ is diagonal) we have:
\begin{align} \label{eq:massdiagrev}
\Mnu = U^* {\rm diag}(m_1, m_2, m_3) U^\dagger \psik
\end{align}
where $\Mnu$ is the neutrino mass matrix, $U$ is the lepton mixing matrix, and $m_i$ are the three neutrino masses. We work in the convention where the neutrino masses absorb the two Majorana phases. Assuming $U$ is of TB form, eq. (\ref{eq:massdiagrev}) gives the general form of $\Mnu$ in terms of the three (complex) neutrino masses $m_1$, $m_2$ and $m_3$:
\begin{align} \label{eq:TBform}
\Mnu &= \left( \begin{array}{ccc} x  & y     &  y    \\
                                  y  & x + z & y - z \\
                                  y  & y - z & x + z \end{array} \right) \psik
\end{align}
\begin{align}
\mbox{with} \lspc x = (2 m_1 + m_2) &/ 3 \shortcomma y = (m_2 - m_1) / 3 \shortcomma z = (m_3 - m_1) / 2 \nekuda
\end{align}
This mass matrix can be described as the most general matrix that is invariant under the action of two unitary symmetric matrices:
\begin{align} \label{eq:TBconds}
\Mnu = S^T \Mnu S \longcomma \Mnu = A_{23}^T \Mnu A_{23} \psik
\end{align}
where $S$ is the one defined in (\ref{eq:irrep}), and $A_{23}$ is a $\mu$-$\tau$ exchange matrix:
\begin{align}
A_{23} = \left( \begin{array}{ccc} 1 & 0 & 0 \\
                                   0 & 0 & 1 \\
                                   0 & 1 & 0 \end{array} \right) \nekuda
\end{align}
Therefore, if $\Me$ is diagonal and $\Mnu$ satisfies eq. (\ref{eq:TBconds}), we get TB mixing.

Let us see how an $A_4$-based theory can produce such a mass matrix. We assume an $A_4$-invariant Lagrangian for leptons, where the neutrino vector $\nu \equiv (\nu_e, \nu_\mu, \nu_\tau)^T$ and the left-handed charged lepton vector $\ell \equiv (e, \mu, \tau)^T$ are triplets of $A_4$, and where the right-handed charged leptons $e^c$, $\mu^c$ and $\tau^c$ transform as 1, 1'' and 1', respectively. This Lagrangian satisfies
\begin{align}
\forall A \in A_4 \longcomma {\cal L}\big[\ell, \nu, \ell^c \big] = {\cal L}\big[D(A) \cdot \ell, D(A) \cdot \nu, \ell^c \cdot D'(A) \big] \psik
\end{align}
where we also use $\ell^c \equiv (e^c, \mu^c, \tau^c)$. $D(A)$ is the $3 \times 3$ matrix associated with the element $A$, namely some multiplication of the matrices $S$ and $T$ from eq. (\ref{eq:irrep}). $D'(A)$ is the same multiplication, substituting the matrix $S$ with the identity matrix and the matrix $T$ with $T^\dagger$ \footnote{This is since the right-handed charged leptons are all singlets, and $S = 1$ for all 1D irreps. In addition, we have $T = \omega$ for $\mu^c$ (the irrep 1'') and  $T = \omega^2$ for $\tau^c$ (the irrep 1'), so we arrange them in a diagonal $3 \times 3$ matrix.}. Let us also assume that, as a result of a particular vacuum alignment, the $A_4$ symmetry is broken down to $G_S$ in the neutrino sector and down to $G_T$ in the charged lepton sector. Then the Lagrangian only satisfies
\begin{align}
{\cal L}\big[\ell, \nu, \ell^c \big] = {\cal L}\big[T \cdot \ell, \nu, \ell^c \cdot T^\dagger \big] = {\cal L}\big[\ell, S \cdot \nu, \ell^c \big] \nekuda
\end{align}
In such a Lagrangian, the charged lepton mass matrix and the neutrino mass matrix (of Majorana type) have to satisfy
\begin{align}
\ell^c \cdot \Me \cdot \ell &= \ell^c \cdot T^\dagger \Me T \cdot \ell \\
\nu^T \cdot \Mnu \cdot \nu &= \nu^T \cdot S^T \Mnu S \cdot \nu \psik
\end{align}
and therefore the mass matrices have to satisfy
\begin{align}
\Me &= T^\dagger \Me T \label{eq:charme} \\
\Mnu &= S^T \Mnu S \nekuda  \label{eq:charmnu}
\end{align}
Eq. (\ref{eq:charme}) characterizes $\Me$ as a diagonal matrix. Given this, eq. (\ref{eq:charmnu}) is exactly one of the conditions we have found for TB mixing (eq. (\ref{eq:TBconds})). Therefore, a model with an added $A_4$ symmetry, with the right symmetry-breaking patterns, can naturally lead to TB mixing. The other condition necessary for TB mixing, $\Mnu = A_{23}^T \Mnu A_{23}$, is not a direct result of the $A_4$ symmetry. In some models it is satisfied by an additional symmetry and in others it is satisfied by imposing some specific limitation on the field content.

Neutrino oscillation experiments also give us information about neutrino mass-squared differences
\begin{align}
\Dsol \equiv \Dt = |m_2|^2 - |m_1|^2 \longcomma \Datm \equiv |\Dth| = ||m_3|^2 - |m_1|^2| \nekuda
\end{align}
$\Dt$ is positive by definition. $\Dth$ is either positive (normal hierarchy) or negative (inverse hierarchy). The analysis \cite{Fogli:2008} results in the following values ($\pm 1 \sigma$) for neutrino mass-squared differences:
\begin{align} \label{eq:massmeas}
\Dsol = 7.67^{+0.16}_{-0.19} \cdot 10^{-5} eV^2 \longcomma \Datm = 2.39^{+0.11}_{-0.08} \cdot 10^{-3} eV^2 \nekuda
\end{align}
We get another small dimensionless parameter,
\begin{align} \label{eq:rvalue}
\rr \equiv \frac{\Dsol}{\Datm} \sim \frac{1}{30} \nekuda
\end{align}
We would expect a natural model to explain the smallness of $\rr$.

In this work we analyze the Altarelli-Feruglio (AF) $A_4$ symmetry model \cite{Altarelli:2005} that follows the guidelines presented above. We show that in order to produce $\rr$ as small as in eq. (\ref{eq:rvalue}), the AF model requires fine tuning, both in its basic form and in its seesaw realization.

\section{Basic model} \label{sec:basic}

In its basic form, the AF model results in the three following neutrino masses:
\begin{align} \label{eq:masses}
m_1 = b + a \longcomma m_2 = a \longcomma m_3 = b - a \nekuda
\end{align}
The numbering of the masses is meaningful, since the model relates between the different masses and the specific mixing angles. The parameters $a$ and $b$ are in general complex, and we define $\ph$ to be the relative phase between them.

First we calculate the two expressions for mass-squared differences:
\begin{equation}
\begin{aligned}
\Dt  = - |b|^2 - 2|a||b|\cph \\
\Dth = - 4|a||b|\cph \nekuda
\end{aligned}
\end{equation}
Since $\Dt$ is positive by definition we must have
\begin{align} \label{eq:uplimit}
\cph < -\frac{1}{2}\left|\frac{b}{a}\right| \sspc ( < 0 ) \nekuda
\end{align}
This makes sense -- we can only have $|b+a| < |a|$ if the angle between $a$ and $b$ is obtuse. Also, since $-1 \le \cph$, we get $|b| \le 2|a|$. This also makes sense -- we cannot have $|b+a| < |a|$ when $|a| \ll |b|$.
Using eq. (\ref{eq:uplimit}), we get
\begin{align}
\Dth = - 4|a||b|\cph > 2|b|^2 \ge 0 \psik
\end{align}
and therefore the basic form of the AF model enforces normal hierarchy \cite{Brahmachari:2008}. Let us investigate it further:
\begin{align}
\rr = \frac{\Dt}{|\Dth|} = \frac{1}{2} \left( \frac{|b|}{2|a|\cph} + 1 \right) \nekuda
\end{align}
We learn that in order to produce a small value for $\rr$, a fine tuned cancelation between the order one parameter $|b|/(2|a|\cph)$ and $1$ has to take place, at the level of $2\rr\sim0.06$. In particular, the parameter $\rr$ that is accounted for by fine tuning is substantially smaller than $\sin\toth$, and comparable to  $\sin \ttth - 1 / \sqrt{2}$ and to $\sin\tot - 1 / \sqrt{3}$, the three small parameters that the model aims to explain by an approximate symmetry.
Notice that higher-order corrections to the three neutrino masses are small (at a level of 5\% of the leading order at most, according to \cite{Altarelli:2005}), and therefore they do not fundamentally affect this conclusion.

\section{Seesaw realization}

Let us do the same analysis for the modified AF model, which accommodates the seesaw mechanism. The model results in the three following neutrino masses:
\begin{align} \label{eq:massesseesaw}
m_1 = \frac{1}{b + a} \longcomma m_2 = \frac{1}{a} \longcomma m_3 = \frac{1}{b - a} \nekuda
\end{align}
We again define $\ph$ as the relative phase between $a$ and $b$. The two expressions for mass-squared differences are:
\begin{equation}
\begin{aligned} \label{eq:massdiff}
\Dt  = \frac{|b|^2 + 2|a||b|\cph}{|a|^2|b+a|^2} \\
\Dth = \frac{4|a||b|\cph}{|b-a|^2|b+a|^2} \nekuda
\end{aligned}
\end{equation}
Since $\Dt$ is positive by definition, we get:
\begin{align} \label{eq:downlimit2}
\cph > -\frac{|b|}{2|a|} \nekuda
\end{align}
This makes sense -- since the absolute values of the masses in the seesaw realization (eq. (\ref{eq:massesseesaw})) are the inverses of the ones in the basic form of the model (eq. (\ref{eq:masses})), we get the exact opposite condition to eq. (\ref{eq:uplimit}). Looking at the expression for $\Dth$ (eq. (\ref{eq:massdiff})) we see that the sign of $\cph$ dictates the type of hierarchy. Since $\cph$ can be either positive or negative (eq. (\ref{eq:downlimit2})), the model can result in either one. For $\rr$, we have
\begin{align}
\rr = \frac{1}{4|\cph|} \cdot \left|1-\frac{b}{a}\right|^2 \cdot \left( \left|\frac{b}{a}\right| + 2\cph \right) \nekuda
\end{align}
Let us see what is necessary in order to get a small $\rr$. We first look at two limits:
\begin{align}
\left|\frac{b}{a}\right| \ll 1 \lspc &\Rightarrow \lspc \rr = \frac{2\cph}{4|\cph|} + O\left(\frac{|b|}{|a|}\right) = \frac{1}{2} + O\left(\frac{|b|}{|a|}\right) \psik \\
\left|\frac{b}{a}\right| \gg 1 \lspc &\Rightarrow \lspc \rr = \frac{1}{4|\cph|} \cdot \left|\frac{b}{a}\right|^2 \cdot \left|\frac{b}{a}\right| > \frac{1}{4}\left(\frac{|b|}{|a|}\right)^3 \gg 1 \nekuda
\end{align}
For the first limit we used the fact that at first order eq. (\ref{eq:downlimit2}) reads $\cph > 0$. In neither case we get the desired range for $\rr$. Thus, the two limiting cases are phenomenologically excluded, and we must have $a$ and $b$ of the same order of magnitude. Given $|a| \sim |b|$, we look for a relation that can make the following expression small:
\begin{align} \label{eq:r23}
\rr = \underbrace{\frac{1}{4|\cph|}}_{\mbox{A}} \cdot \underbrace{\left|1-\frac{b}{a}\right|^2}_{\mbox{B}} \cdot \underbrace{\left( \left|\frac{b}{a}\right| + 2\cph \right)}_{\mbox{C}}
\end{align}

\begin{figure}[h]
  \centering
  \includegraphics[width=0.75\textwidth]{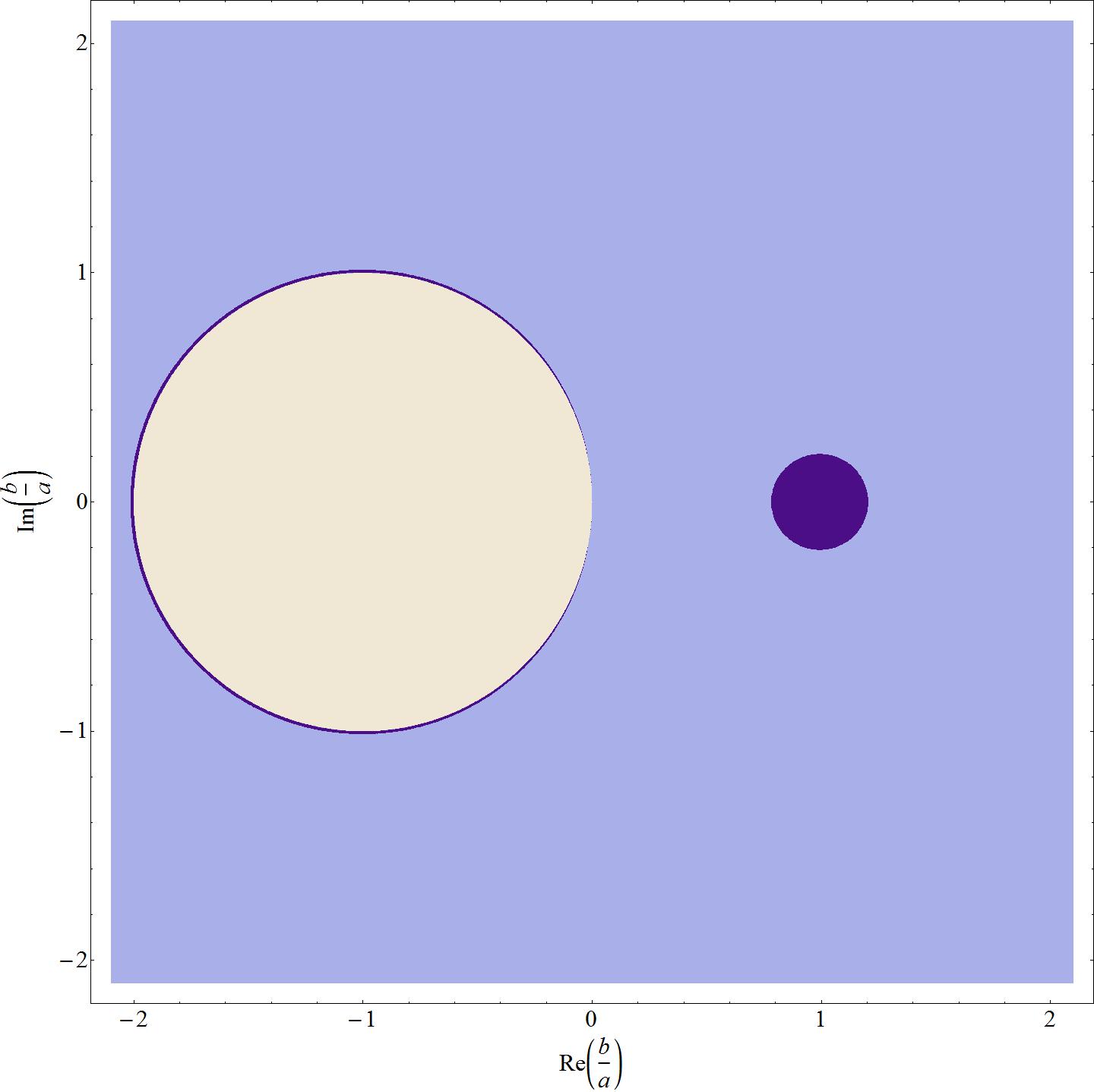}
  \caption{Different ranges of $\rr$ as a function of the complex variable b/a. The yellow area is forbidden ($\rr < 0$), the light purple area provides $1/30 < \rr$ and the dark purple areas provide $0 < \rr < 1/30$.}
  \label{fig:areas}
\end{figure}

A is not smaller than 1/4. So we need either B or C to be "surprisingly" small. B would be small if $a$ and $b$ are very similar in size and in phase. C would be small if  $2\cph \sim -|b/a|$. Since $\rr$ only depends on the complex number $b/a$ (with $\ph$ as its phase), we can draw it on a 2D plane. In figure \ref{fig:areas} we see what values of $b/a$ provide a small enough $\rr$. We can identify the two areas that we mentioned: $b/a \sim 1$ (the full circle on the right) and $2\cph \sim -|b/a|$ (the empty circle on the left). Overall, these areas are very small, so some fine tuning is needed for this model too. Again, higher-order corrections to the three neutrino masses are small (at a level of 5\% of the leading order at most, according to \cite{Altarelli:2005}), and therefore they do not fundamentally affect this conclusion.

In ref. \cite{Altarelli:2005} there is a brief analysis of the possible hierarchies allowed by the seesaw realization of the AF model. Two examples are given to demonstrate strong hierarchy (small $\rr$): $b \sim a$ and $b \sim -2 a$. Indeed, these are two "dark purple" areas in our graph. Let us calculate the level of fine tuning around these points, assuming real $a$ and $b$ for simplicity. For real parameters, eq. (\ref{eq:r23}) reduces to
\begin{align} \label{eq:realr23}
\rr = \frac{1}{4} \left(1-\frac{b}{a}\right)^2 \left( 2 + \frac{b}{a} \right) {\rm sgn}  \left( \frac{b}{a} \right) \nekuda
\end{align}
For the area $b \sim a$ we define our small parameter $\e$ by:
\begin{align}
b = a(1 + \e) \nekuda
\end{align}
At leading order in $\e$ eq. (\ref{eq:realr23}) gives
\begin{align}
\rr = \frac{3}{4} \e^2 \nekuda
\end{align}
Notice that the area $b \sim a$ (with $\cph > 0$) leads to normal hierarchy. For normal hierarchy, the relevant flavor parameter is $m_2 / m_3 \approx \sqrt{\rr}$, consistent with $\e \sim \sqrt{\rr}$. Demanding $\rr < 1/30$, we get the following constraint on $\e$:
\begin{align}
-0.2 < \e < 0.2 \nekuda
\end{align}
Indeed, the radius of the full circle in figure \ref{fig:areas} is about 0.2. For the area $b \sim -2 a$ we define our small parameter $\e$ by:
\begin{align}
b = -2a(1 + \e) \nekuda
\end{align}
At leading order in $\e$ eq. (\ref{eq:realr23}) gives
\begin{align}
\rr = \frac{9}{2} \e \nekuda
\end{align}
Notice that the area $b \sim -2 a$ (with $\cph < 0$) leads to inverse hierarchy. For inverse hierarchy, the relevant flavor parameter is $\Dt / m_2^2 \simeq \rr$, consistent with $\e \sim \rr$. Demanding $\rr < 1/30$, we get the following constraint on $\e$:
\begin{align}
0 < \e < \frac{1}{135} \nekuda
\end{align}
Indeed, the width of the empty circle in figure \ref{fig:areas} is about 2/135.

Doing this analysis, it might seem like the level of fine tuning is not so high (having $0.8a < b < 1.2a$ is enough). But this is because it assumes real $a$ and $b$. Figure \ref{fig:areas} shows us that assuming real $a$ and $b$ (namely, keeping to the $x$ axis) increases the probability to get a small $\rr$. Taking into account the relative phase between $a$ and $b$ (as we should), namely looking at the whole complex space, we get much smaller probability for such a small $\rr$. In fact, the total "dark purple" area in the graph is about 0.2. So if we assume, for example, that $b/a$ is a random complex number with an absolute value between 1/3 and 3, we get a probability of about 1/120 to get $\rr < 1/30$.

\section{Conclusions} \label{sec:concs}

We examined the two realizations of the AF model in terms of the level of fine tuning required to explain the smallness of $\rr$. For each one of the realizations we calculated the theoretical prediction for the two mass-squared differences. We saw that the basic form of the AF model enforces normal hierarchy and that its seesaw realization allows both hierarchies. For the basic model, the strong hierarchy (small $\rr$) indicates that order-one real parameters should cancel each other at a level of $2\rr$. For the seesaw realization, the strong hierarchy indicates that an order-one complex parameter is fine-tuned to a small area in the complex plane. The simplification of making this parameter real (as was done in ref. \cite{Brahmachari:2008}) does not give us the full picture in terms of fine tuning, since only a very small part of the rest of the complex plane leads to a small enough $\rr$. A similar conclusion was drawn by numerical means in ref. \cite{Barry:2010}.

The AF model leads to some finite value for $\rr$ in the symmetry limit. Since all relevant parameters are of order one, this finite value is expected to be order one and we get a fine-tuning problem. In other words, the AF model does not give a natural explanation for the smallness of $\rr$. (A similar problem occurs in models with Abelian flavor symmetries \cite{Amitai:2012}.) A model that gives such an explanation would have to have $\rr = 0$ in the symmetry limit, and then get some small corrections due to symmetry breaking. In fact, the current measured values satisfy $\rr \sim \sin^2 \toth$. Therefore, assuming that $\toth$ is a result of corrections at first order of some symmetry-breaking parameter, like in the AF model, $\rr$ should not only vanish in the symmetry limit but ideally arise only at second order in this parameter.

\section*{Acknowledgments}

I thank Yossi Nir for useful discussions and assistance with the draft. The research of SA is supported by the Israel Science Foundation.

\bibliography{FineTuningBib}
\bibliographystyle{utphys}

\end{document}